\documentclass[fleqn, 12pt]{article}

\begin{document}

\title{Short range gravitational fields:\\
Rise and fall of the fifth force}

\author{Luigi Foschini\\
{\small CNR Institute FISBAT, Via Gobetti 101, I-40129, Bologna (Italy)}\\ 
{\small e-mail:  L.Foschini@fisbat.bo.cnr.it}}

\maketitle
          
\begin{abstract}
During the 80's, some experiments and the repetitions of old ones, lead to the 
hypothesis of a fifth force. Nevertheless, a more accurate research was not able 
to confirm this hypothesis. This article wants to go over again the most important 
steps of the event.
\end{abstract}

\vskip 24pt

\noindent PACS-01.65.+g   History of science.\\
PACS-04.80.Cc  Experimental tests of gravitational theories.

\vskip 48pt

\begin{center}
CNR - Bologna Research Area Library\\
cat. 8707 Reports - coll. 126-4-113,1997
\end{center}

\newpage
\section{Introduction}
In the study of dynamics, the concept of mass is introduced relatively to the laws of 
motion; it is therefore called \emph{inertial mass} ($m_{i}$). It is a universal 
concept and it concerns with all kind of matter, whether it be a proton or an apple.

Studying gravitation, another kind of mass is introduced instead, which is called 
\emph{gravitational} ($m_{g}$), necessary to characterize the intensity of this 
interaction. It is then posssible to divide the gravitational mass in active and 
passive: this depends on whether it undergoes or generates the gravitational field.

If gravitation is a universal property too, just like the laws of dynamics, then the 
ratio:

\begin{equation}
k=\frac{m_{g}}{m_{i}}
\label{e:ratio}
\end{equation}

\noindent does not depend  on the type of matter and, choosing the units of measure 
opportunely, one will obtain $k=1$. This choice is implicitly made when the universal 
gravitational constant $G$ is defined, therefore the constance of $k$ is equal to that 
of $G$.

This is the so called \emph{principle of weak equivalence}, to be distinguished from 
other forms of the principle of equivalence (medium and strong) which regards other 
aspects of the gravitational theory, more linked with the theory of general relativity 
(see \cite{Ciufolini}). 

The most famous experiment to verify the principle of equivalence is, without doubt, 
that made by Galileo Galilei at the beginning of XVII century. Other experiments 
followed, by Newton, Huygens, Bessel just to quote some of them. An important step was 
realized by von E\"{o}tv\"{o}s in 1889 and in 1922, obtaining an accuracy of 
$5\cdot 10^{-8}$ \cite{Eotvos}. Later on, in 1964 Roll, Krotov and Dicke reached
$10^{-11}$, using the gravitational field of the Sun \cite{Roll}. Again, other 
experiments were made, especially on an interplanetary scale, which confirmed the 
validity of the principle of equivalence (see \cite{Ciufolini}, \cite{Will}). 
On a geophysical scale instead, one had stopped at von E\"{o}tv\"{o}s' results of 
1922.

\section{The first skirmishes}
In 1981, an article by Stacey, Tuck, Holding, Maher and Morris was published 
dealing with some anomalous measures of the universal gravitational constant $G$, 
made in a mine site in Queensland, Australia \cite{Stacey1}. In this article and 
in others, Stacey's group underlines a value of $G$, obtained through the 
measures of the acceleration of gravity $g$ in mines, also considering many 
corrective factors \cite{Holding}, \cite{Stacey1}. Therefore they think that the
explanation of this anomaly lies in the presence of a short-range potential, of the 
Yukawa type, that overlaps the newtonian potential:

\begin{equation}
U=-G_{\infty}\frac{m}{r}(1+\alpha e^{-\frac{r}{\lambda}})
\label{e:fish}
\end{equation}

\noindent where $\alpha$ and $\lambda$ are respectively the intensity and the 
action range of the Yukawa type potential and $G_{\infty}$ is the value of $G$ 
going to infinite, where Eq.~(\ref{e:fish}) is to be reduced to the traditional 
Newton equation. The principle of equivalence has as consequence the constance 
of $G$ in time and space, therefore a variation on geophysical scale would lead 
to the negation of the very principle. As far as Eq.~(\ref{e:fish}), for $r<<\lambda$ 
one would have: 

\begin{equation}
G(r)\approx G_{\infty}(1+\alpha-\frac{\alpha r^{2}}{2\lambda^{2}}+\ldots)
\label{}
\end{equation}

Therefore, a new type of interaction, which should be added to the other four known,
that is the gravitational, the electromagnetic and the strong and weak nuclear 
interaction (actually they should be three, because the electromagnetic and the 
weak nuclear were unified at the end of the 70's by A. Salam, S. Glashow, 
S. Weinberg and the experimental test was achieved thanks to C. Rubbia at the
beginning of the 80's).

To these achievements, an article by Fischbach \emph{et al.} \cite{Fischbach1} 
was to be added soon. After a reanalysis of the experiments lead by von 
E\"{o}tv\"{o}s on the principle of equivalence between the inertial and 
gravitational mass \cite{Eotvos}, Fischbach \emph{et al.} find out a similar 
course and attributes it to a short-range interaction depending upon the material. 
From the authors' viewpoint this interaction had not been observed in the most precise 
experiment by Roll, Krotov and Dicke \cite{Roll}, for these are referred to the 
Sun and to a wide distance that makes the potential effects of this new interaction 
negligible. As a matter of facts, the two parameters $\alpha$ and $\lambda$ which 
appear in Eq.~(\ref{e:fish}) and which would be respectively the intensity and 
action range of the fifth interaction, have these values: $|\alpha| \approx
10^{-2}\div 10^{-3}$ and $1\leq \lambda \leq 10^{4}$ m.

\section{A nest of hornets}
At this point a nest of hornets was stirred up: a pair of comments, meant to 
stress some thoughtlesses, were directed towards Fischbach's group, in particular to 
notify a wrong sign for $\alpha$ \cite{Keyser}, \cite{Thodberg}: Fischbach \emph{et
al.} \cite{Fischbach1} spoke about a repulsive force though their analysis showed 
an attractive one. To these objections the authors answered that the 
sign change did not invalidate the global reasoning \cite{Fischbach2}. However, one 
has to mark that the sign $\alpha$ will change several times, as well as the limits 
of $\lambda$, insomuch that the evolution of these two parameters results more 
complicated than a hiccupping snake's route.

Stacey's group shows up to help Fischbach's group with a new analysis of the anomalies
took in the mines \cite{Holding}, \cite{Stacey2}, \cite{Stacey3} and three other 
experiments. The first one was caried out by Thieberger \cite{Thieberger}, 
realizing an experiment 
based upon a hollow copper sphere bathed into water and free to move about: this 
should make it possible to measure the differences of acceleration between liquid 
and solid. Thieberger put the instrument on an precipice on the 
Hudson River, New Jersey: this is already something that puzzles us because it did 
not seem a good idea to place oneself near a river, therefore near to a flux of 
variable mass, to make an extremely accurate gravitational experiment. Anyway, 
Thieberger himself, in a footnote at the end of the article \cite{Thieberger}, 
reported that M. L. Good made him notice that the Coriolis' acceleration was not 
negligible in those circumstances and taking this under consideration reduces the 
deviation of the measured $G$. A similar experiment made in Italy by Bizzeti \emph{et
al.} \cite{Bizzeti} did not point out notable variations.

The second one is a group leaded by Boynton, that repeats von E\"{o}tv\"{o}s'
experiment though in two different places; first on a precipe on the North Cascades, 
near to Index, Washington, then in a building of the department in Seattle
\cite{Boynton}. Some deviations from the value of $G$ are noticed, though the 
disturbancies may be held responsible again.

A third experiment is made by Eckhardt \emph{et al.} \cite{Eckhardt1}: it consists 
in the measuring of the gravitational acceleration at ground level and at various 
distances from the ground, by climbing up a telecasting tower of 600 m. The 
experiment showed immediately its very limits: in a commentary, Bartlett and Tew 
proved that the effect of the surrounding ground had been underestimated
\cite{Bartlett}. Eckhardt group's questioned the feasibility of Bartlett 
and Tew's analitical method though not the possibility of a presence of ground 
effects \cite{Eckhardt2} and, in a later and more accurate data analysis, the 
anomaly disappears \cite{Jekeli}.

In that period, a number of experiments were made, but they all gave negative results: 
on towers \cite{Speake}, \cite{Thomas1}; the typical experiment of the free fall from 
a tower \cite{Niebauer}; on a wider scale with submarine measures \cite{Zumberge} or 
also near a channel's lock \cite{Bennett}.

\section{Is the Earth a perfect sphere?}
There are other very important experiments: in 1990, Thomas and Vogel \cite{Thomas2} 
wrote a very witty and interesting article disheartening many ideas of Stacey's group. 
The first, and perhaps the most important point, is that in the 
newtonian gravitational law, Earth is considered to be a homogeneous sphere, 
uniform and not--spinning. The famous formula:

\begin{equation}
U(r)=-G\frac{m}{r}
\label{e:newton}
\end{equation}

\noindent is referrred to a perfect sphere. Nevertheless, reality is a little bit 
different and the Earth is lightly crushed at the poles, being therefore a spheroid. 
What is more, it is not even stationary, homogeneous and uniform. One can remeber 
that the Earth is spinning, introducing a centrifugal potential: 

\begin{equation}
U_{c}(r, \varphi)=\frac{1}{2}\omega^{2}r^{2}\cos^{2}\varphi
\label{e:centrif}
\end{equation}

\noindent where $\omega$ represents the angular velocity and $\varphi$ the latitude. 
The crushing of the poles instead produces a deformation of the gravitational field, 
expressible with a series of spherical harmonics:

\begin{equation}
U(r, \theta, \phi)=G\frac{m}{r}\sum_{n=0}^{\infty}\left(\frac{a}{r}\right)^{2}
\sum_{m=0}^{n}(A_{n}^{m}\cos m\phi+B_{n}^{m}\sin m\phi) P_{n}^{m}(\theta)
\label{e:sph}
\end{equation}

\noindent where $a$ is the equatorial radius, $\phi$ is the longitude, $\theta$ is the 
colatitude and $P_{n}^{m}(\theta)$ are Legendre's normalized functions in degree $n$ 
and in order $m$. Eq.~(\ref{e:sph}) describes the gravitational potential of the Earth 
as the sum of the potentials of endless ideal masses (monopoles, dipoles, \ldots), 
centred in the origin and with a statistic weight, due to the coefficients $A_{n}^{m}$ 
and $B_{n}^{m}$. Eq.~(\ref{e:sph}) can be considerably semplified taking under 
consideration various symmetries and geometries (see \cite{Bertotti}, \cite{Blakely}); 
therefore at the end of the process one can write:

\begin{equation}
U_{s}(r, \varphi)=-G\frac{m}{r}[1-J_{2}\left(\frac{a}{r}
\right)^{2}\frac{(3\sin^{2}\varphi-1)}{2}]
\label{e:spsem}
\end{equation}

\noindent where $J_{2}=1.082626\cdot 10^{-3}$ is the (dimensionless) coefficient of 
ellipticity of the Earth spheroid. Eq.~(\ref{e:spsem}) says that, in terms of spherical 
armonics, the contribution of the quadrupoles is the most important one, after the 
newtonian potential, which is the first term of the series.

Then, the Earth gravitational potential can be expressed, with proper approximation, 
as:

\begin{equation}
U=U_{s}+U_{c}
\label{e:pot}
\end{equation}

Eq.~(\ref{e:spsem})is very important, because it shows that the sole fact that 
the Earth is a spheroid, and not a sphere, involves an additional term of about 
$10^{-3}$, very near to that of $\alpha$, which shows the intensity of the Yukawa 
type potential.

There are further factors of correction to be notified. For example the correction 
for free air refers to anomalies introduced by observations above sea level and it 
can be given with the formula (see \cite{Blakely}):

\begin{equation}
g_{a}=-0.3086\cdot 10^{-5}\cdot h
\label{e:air}
\end{equation}

\noindent where $h$ is the elevation above sea level. It is to be noticed that 
being $g/h$ expressed in s$^{-2}$, Eq.~(\ref{e:air}) is valid both in SI ($g_{a}$ in 
m/s$^{2}$; $h$ in m), and CGS ($g_{a}$ in cm/s$^{2}$; $h$ in cm).

Bouguer's correction instead, takes under consideration the additional masses above 
sea level, considering them as flat, endless, homogenuous slabs, equal in thickness 
to the altitude of the point of observation. Supposing that the mean typical 
density of the Earth's crust is equal to $2760$ kg/m$^{3}$, Bouguer's corrections 
turns out to be (see \cite{Blakely}):

\begin{equation}
g_{B}=0.1119\cdot 10^{-5}\cdot h
\label{e:bou}
\end{equation}

What has just been said about unity of measure is valid here as well. Using this 
anomaly, along with the study of seismic waves propagation, it was possible to 
identify the Chicxulub crater, where the catastrophic impact, leading to dinosaurs 
extinction 65 milions of years ago, probably took place \cite{Hildebrand}. In this 
case, the comparison with seismic data allows to consider this research as an 
indirect evidence for the classical theory validity.

Going back to Thomas and Vogel's article, they report a series of measure of gravity, 
taken near the sites for nuclear American tests in Nevada \cite{Thomas2}. In this 
case the $G$ value has undergone deviations up to a 4\% rate, against the 1\% in 
Stacey's measures. However, from the seismic analysis of the nuclear explosion, it 
has been possible to discover a reflection barrier at 10 km of depth that makes one 
suppose the existance of high density material, such to generate anomalous 
potential gradients. If, at this point, one analyses Stacey group's data, it will 
grow extremely probable that the anomaly hypotesis is due to the presence of some 
high density material, which is quite common in a mine. Other experiments, made in 
Greenland gave similar results \cite{Ander}: the observed anomalies were due to 
intrusion in the ice of material at a high density.

There is still a kind of experiment on this scale left, those made by using the 
variation of the mass in pumped--storage reservoires for hydroelectric power plants. 
In 1989, M\"{u}ller \emph{et al.} \cite{Muller} traced no significant anomaly, 
after having taken some measure near Lake Hornberg in Germany. In 1997 instead, 
Achilli \emph{et al.} \cite{Achilli} obtained a positive result near Lake Brasimone. 
Making a comparison of the two experiments, one notices some factors which may be 
crucial: differing from Lake Brasimone, Lake Hornberg has an asphalted ground, 
therefore water infiltrations are extremely reduced. What is more 
M\"{u}ller's group used two gravimeters above and under the lake, 
while Achilli's group used one only, under the lake. The experimental error in 
M\"{u}ller \emph{et al.} is nearly totally due to the calibration of 
the two gravimeters ($0.25\pm0.4$ \%); in the case of 
Achilli \emph{et al.} instead, to obtain a very reduced error (0.1\%) a particular 
device was used, because with the usual method the error was to high (1\%).

\section{How to get rid of local anomalies}
Taken into consideration what we said, one may well see that experiments on a 
geophysical scale are far too dependent on the anomalies of density present in the 
Earth crust. The possible alternatives are two: the first is to reproduce on
laboratory scale the Roll, Krotov, and Dicke's experiment \cite{Roll}; the 
second is to make experiments in space, in situations where the gravitational 
perturbations would be reduced as much as possible. 

The first was made in 1990 \cite{Adelberger} and repeated, with some improvements in
1994 \cite{Su}: it was called the E\"{o}tWash experiment. This is, perhaps, the most 
precise experiment up to now realized on laboratory scale in the terrrestrial field. 
Various contrivances were used to compensate the possible anomalies and, at the end of 
measurement, the comparison between the inertial and the gravitational mass for two 
samples of copper and beryllium resulted as \cite{Adelberger}:

\begin{equation}
\frac{m_{i}}{m_{g}}(\mathrm{Cu})-\frac{m_{i}}{m_{g}}(\mathrm{Be})=
(0.2\pm 1.0)\cdot 10^{-11}
\label{e:eotwash}
\end{equation}

\noindent the same order of precision obtained by Roll, Krotov and Dicke \cite{Roll}. 
Moreover, Su \emph{et al.} \cite{Su} reached about $10^{-12}$ with several other 
samples. Therefore they did not find any fifth force, 
forcing its possible intensity to values still more negligible.

The second possibility, that of experiments in space has not been brought about yet. 
Many proposals have been made and one counts much on the future \emph{International
Space Station}, as research laboratory for experimental physics on gravitational 
fields \cite{Spallicci}. Another experiment called \emph{Galileo Galilei}"
was accuratedly elaborated: it is to be performed in space and it should reach a 
precision of $10^{-17}$ \cite{Nobili}.

\section{Conclusions?}
In March 1992, Fischbach and Talmadge wrote an article to see what the situation was 
\cite{Fischbach3}. Their conclusion was that, even though there lacked any evidence 
for the fifth force, the anomalies found by Thieberger \cite{Thieberger} and the 
Boynton's group \cite{Boynton} were to be explained still. These conclusions are 
\emph{definitively diplomatic} for they do not underline that it is 
possible to explain those anomalies by a wrong evaluation of perturbations. As 
Adelberger \emph{et al.} \cite{Adelberger} stress, it was never possible to 
reproduce these anomalies, not even by the researchers who showed them. 

Nevertheless, one has to point out that this episode has spurred a quantity of 
experiments, which allowed to improve our knowledge of the gravitational 
field on a small scale.

\section{Acknowledgements}
A sincere thank to Federico Palmonari and Anna Nobili for the challenging 
discussions on the theme. Author wishes to thank also Simona Baldoni for valuable help
in english translation.

\end{document}